\documentclass[sigplan]{acmart}

\renewcommand\footnotetextcopyrightpermission[1]{} 
\usepackage{epsfig,endnotes,url,color,subfigure}
\usepackage{array}
\usepackage{multirow}
\usepackage{xspace}
\usepackage{verbatim}
\usepackage{wrapfig}
\usepackage[english]{babel}
\usepackage{amsmath}
\usepackage{caption}
\usepackage{balance}
\usepackage[ruled, vlined]{algorithm2e}

\usepackage{hyperref}
\hypersetup{pdftex,colorlinks=true,allcolors=blue}
\usepackage{hypcap}
\usepackage{xr}

\newcommand{\projecttitle}{\textsc{StreamApprox}\xspace}
\newcommand{\myfontsize}{\fontsize{8}{9}\selectfont}
\newcommand{\commentfontsize}{\fontsize{7}{8}\selectfont}

\makeatletter
\def\@copyrightspace{\relax}
\makeatother
\setcopyright{none}

\newcommand{\myparagraph}[1]{\smallskip \noindent{\bf {#1}.}}

\newcommand{\out}[1] {}

\newcounter{codeLineCntr}

\reversemarginpar
\newif\ifnotes
\notestrue

\newcommand{\punt}[1]{}

\renewcommand{\eqref}[1]{Equation~(\ref{eq:#1})}

\newcommand{\proc}[1]{\ifmmode\mbox{\textsc{#1}}\else\textsc{#1}\fi}

  \newcommand{\func}[1]{\ifmmode\mathrm{#1}\else\textrm{#1}fi} %

\newcounter{remark}[section]

\usepackage[inline]{enumitem}
\setlist{noitemsep,topsep=0pt,parsep=0pt,partopsep=0pt}

\usepackage[subtle]{savetrees}

\usepackage{setspace}
\usepackage[margin=5pt,font={stretch=0.9}]{caption}

\begin{document}
\title{Approximate Stream Analytics in \\Apache Flink and Apache Spark Streaming}
\subtitle{An Online Algorithm for Distributed Approximate Stream Analytics}

\author{Do Le Quoc$^1$,  Ruichuan Chen$^2$, Pramod Bhatotia$^{3}$,\\ Christof Fetzer$^1$, Volker Hilt$^2$, Thorsten Strufe$^1$}
\affiliation{	$^1$TU Dresden, $^2$Nokia Bell Labs, $^3$University of Edinburgh and Alan Turing Institute}

\begin{abstract}
Approximate computing aims for efficient execution of workflows where an approximate output is sufficient instead of the exact output. The idea behind approximate computing is to compute over a representative sample instead of  the entire input dataset. Thus, approximate computing --- based on the chosen sample size --- can make a systematic trade-off between the output accuracy and computation efficiency.

Unfortunately, the state-of-the-art systems for approximate computing primarily target batch analytics, where the input data remains unchanged during the course of sampling. Thus, they are not well-suited for stream analytics. This motivated the design of \projecttitle --- a stream analytics system for approximate computing. To realize this idea, we designed an online stratified reservoir sampling algorithm to produce approximate output with rigorous error bounds.  Importantly, our proposed algorithm is generic and can be applied to two prominent types of stream processing systems: (1) batched stream processing such as Apache Spark Streaming, and (2) pipelined stream processing such as Apache Flink.

To showcase the effectiveness of our algorithm, we implemented \projecttitle as a fully functional prototype based on Apache Spark Streaming and Apache Flink. We evaluated \projecttitle using a set of microbenchmarks and real-world case studies. Our results show that Spark- and Flink-based \projecttitle systems achieve a speedup of $1.15\times$---$3\times$ compared to the respective native Spark Streaming and Flink executions, with varying sampling fraction of $80\%$ to $10\%$.  Furthermore, we have also implemented an improved baseline in addition to the native execution baseline --- a Spark-based approximate computing system leveraging the existing sampling modules in Apache Spark. Compared to the improved baseline, our results show that \projecttitle achieves a speedup $1.1\times$---$2.4\times$ while maintaining the same accuracy level.
This technical report is an extended version of our conference publication~\cite{streamapprox-middleware}.


 


\end{abstract}

\maketitle

\section{Introduction}
\label{sec:introduction}

Stream analytics systems are extensively used in the context of modern online services to transform continuously arriving raw data streams into useful insights~\cite{flink,  naiad, d-streams}. These systems target low-latency execution environments with strict service-level agreements (SLAs) for processing the input data streams.

In the current deployments, the low-latency requirement is usually achieved by employing more computing resources and parallelizing the application logic over the distributed infrastructure. Since most stream processing systems adopt a data-parallel programming model~\cite{mapreduce}, almost linear scalability can be achieved with increased computing resources.

However, this scalability comes at the cost of ineffective utilization of computing resources and reduced throughput of the system. Moreover, in some cases, processing the entire input data stream would require more than the available computing resources to meet the desired latency/throughput guarantees.

To strike a balance between the two desirable, but contradictory design requirements --- low latency and efficient utilization of computing resources --- there is a surge of {\em approximate computing} paradigm that explores a novel design point to resolve this tension. In particular, approximate computing is based on the observation that many data analytics jobs are amenable to an approximate rather than the exact output~\cite{approx-ex-1,approx-ex-2}. For such workflows, it is possible to trade the output accuracy by computing over a subset instead of the entire data stream. Since computing over a subset of input requires less time and computing resources, approximate computing can achieve desirable latency and computing resource utilization. 

To design an approximate computing system for stream analytics, we need to address the following three important design challenges: Firstly, we need an online sampling algorithm that can perform ``on-the-fly'' sampling on the input data stream. Secondly, since the input data stream usually consists of sub-streams carrying data items with disparate population distributions, we need the online sampling algorithm to have a ``stratification'' support to ensure that all sub-streams (strata) are considered fairly, i.e., the final sample has a representative sub-sample from each distinct sub-stream (stratum). Finally, we need an error-estimation mechanism to interpret the output (in)accuracy using an error bound or confidence interval.

Unfortunately, the advancements in approximate computing are primarily geared towards batch analytics~\cite{BlinkDB, approxhadoop, quickr-sigmod}, where the input data remains unchanged during the course of sampling (see $\S$\ref{sec:related} for details). In particular, these systems rely on pre-computing a set of samples on the static database, and take an appropriate sample for the query execution based on the user's requirements (aka query execution budget). Therefore, the state-of-the-art systems cannot be deployed in the context of stream processing, where the new data continuously arrives as an unbounded stream.

As an alternative, we could in principle {\em repurpose} the available sampling mechanisms in Apache Spark  (primarily available for machine learning in the MLib library~\cite{mlib}) to build an approximate computing system for stream analytics. In fact, as a starting point, we designed and implemented an approximate computing system for stream processing in Apache Spark based on the available sampling mechanisms.
Unfortunately, as we will show later, Spark's stratified sampling algorithm suffers from three key limitations for approximate computing, which we address in our work (see $\S$\ref{sec:implementation} for details). First, Spark's stratified sampling algorithm operates in a ``batch'' fashion, i.e., all data items are first collected in a batch as Resilient Distributed Datasets (RDDs)~\cite{spark-nsdi-2012}, and thereafter,  the actual sampling is carried out on the RDDs. Second,  it does not handle the case where the arrival rate of sub-streams changes over time because it requires a pre-defined sampling fraction for each stratum. Lastly, the stratified sampling algorithm implemented in Spark requires synchronization among workers for the expensive join operation, which imposes a significant latency overhead.

To address these limitations, we designed an {\em online stratified reservoir sampling algorithm} for stream analytics. Unlike existing Spark-based systems, we perform the sampling process ``on-the-fly'' to reduce the latency as well as the overheads associated in the process of forming RDDs.  Importantly, our algorithm {\em generalizes to two prominent types of stream processing models}: (1) batched stream processing employed by Apache Spark Streaming~\cite{spark-streaming}, and (2) pipelined stream processing  employed by Apache Flink~\cite{flink}.

More specifically, our sampling algorithm makes use of two techniques: reservoir sampling and stratified sampling. We perform reservoir sampling for each sub-stream by creating a fixed-size reservoir per stratum.  Thereafter, we assign weights to all strata respecting  their respective arrival rates to preserve the statistical quality of the original data stream. The proposed sampling algorithm naturally adapts to varying arrival rates of sub-streams, and requires no synchronization among workers (see \S\ref{sec:design}).



\begin{figure}[t]
\centering
\includegraphics[scale=0.33]{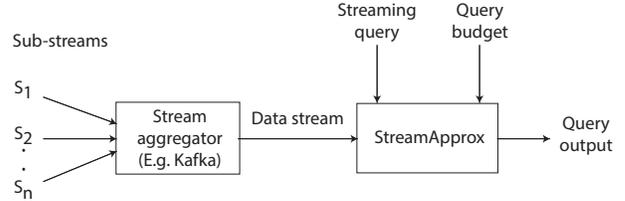}

\caption{System overview}

\label{fig:sysOverview}
\end{figure}

Based on the proposed sampling algorithm, we designed \projecttitle, an approximate computing system for stream analytics (see Figure~\ref{fig:sysOverview}). \projecttitle provides an interface for users to specify streaming queries and their execution budgets. The query execution budget can be specified in the form of latency guarantees or available computing resources.  Based on the query budget, \projecttitle provides an adaptive execution mechanism to make a systematic trade-off between the output accuracy and computation efficiency. In particular, \projecttitle employs the proposed sampling algorithm to select a sample size based on the query budget, and executes the streaming query on the selected sample.  Finally, \projecttitle provides a confidence metric on the output accuracy via rigorous error bounds.  The error bound gives a measure of accuracy trade-off on the result due to the approximation.

%
%
%
%

We implemented \projecttitle based on Apache Spark Streaming~\cite{spark-streaming} and Apache Flink~\cite{flink}, and evaluate its effectiveness via various microbenchmarks. Furthermore, we also report our experiences on applying \projecttitle to two real-world case studies. 
Our evaluation shows that Spark- and Flink-based \projecttitle achieves a significant speedup of $1.15\times$ to $3\times$ over the native Spark Streaming and Flink executions, with varying sampling fraction of $80\%$ to $10\%$, respectively.

In addition, for a fair comparison, we have also implemented an approximate computing system leveraging the sampling modules already available in Apache Spark's MLib library (in addition to the native execution comparison). Our evaluation shows that, for the same accuracy level, the throughput of Spark-based \projecttitle is roughly $1.1\times$---$2.4\times$ higher than the Spark-based approximate computing system for stream analytics.



\vspace{5mm}
To summarize, we make the following main contributions.
\begin{itemize}
\item We propose the online adaptive stratified reservoir sampling (OASRS) algorithm that preserves the statistical quality of the input data stream, and is resistant to the fluctuation in the arrival rates of strata. Our proposed algorithm is generic and can be applied to the two prominent stream processing models: batched and pipelined stream processing models.

\item We extend our algorithm for distributed execution. The OASRS algorithm can be parallelized naturally without requiring any form of synchronization among distributed workers.


\item We provide a confidence metric on the output accuracy using an error bound or confidence interval.  This gives a measure of accuracy trade-off on the result due to the approximation.

\item Finally, we have implemented the proposed algorithm and mechanisms based on Apache Spark Streaming and Apache Flink. We have extensively evaluated the system using a series of microbenchmarks and real-world case studies.

\end{itemize}

\projecttitle's codebase with the full experimental evaluation setup is publicly available: \href{https://streamapprox.github.io/}{https://streamapprox.github.io/}.

\section{Overview and Background}
\label{sec:overview}

This section gives an overview of \projecttitle,  its computational model, and the design assumptions. Lastly, we conclude this section with a brief  background on the technical building blocks.

\subsection{System Overview}
\label{subsec:overview}

\projecttitle is designed for real-time stream analytics. Figure~\ref{fig:sysOverview} presents the high-level architecture of \projecttitle.  The input data stream usually consists of data items arriving from diverse sources.  The data items from each source form a \emph{sub-stream}.  We make use of a stream aggregator (e.g., Apache Kafka~\cite{kafka}) to combine the incoming data items from disjoint sub-streams. \projecttitle then takes this combined stream as the input for data analytics.

We facilitate data analytics on the input stream by providing an interface for users to specify the streaming query and its corresponding query budget. The query budget can be in the form of expected latency/throughput guarantees, available computing resources, or the accuracy level of query results.

\projecttitle ensures that the input stream is processed within the specified query budget.  To achieve this goal, we make use of approximate computing by processing only a subset of data items from the input stream, and produce an approximate output with rigorous error bounds. In particular, \projecttitle designs a parallelizable online sampling technique to select and process a subset of data items, where the sample size can be determined based on the query budget.

\subsection{Computational Model} 
\label{subsec:system-model}

The state-of-the-art distributed stream processing systems can be classified in two prominent categories:  {\em (i)} batched stream processing model, and {\em (ii)}  pipelined stream processing model. These systems offer three main advantages: (a) efficient fault tolerance, (b) ``exactly-once'' semantics, and (c) unified programming model for both batch and stream analytics. {\em Our proposed algorithm for approximate computing is generalizable to both stream processing models, and preserves their advantages.}

\myparagraph{Batched stream processing model} In this computational model, an input data stream is divided into small batches using a predefined batch interval, and each such batch is processed via a distributed data-parallel job. Apache Spark Streaming~\cite{spark-streaming} adopted this model to process input data streams. While this model is widely used for many applications, it cannot adapt to the cases where low-latency is critical since this model waits for the batching to complete before processing the batch. Sampling the input data stream in a continuous ``on-the-fly'' fashion can be challenging to address in this computational model. However, \projecttitle overcomes this challenge by  performing sampling operations before the batches are formed.

\myparagraph{Pipelined stream processing model} In contrast to the batched stream processing model, the pipelined model streams each data item to the next operator as soon as the item is ready to be processed without forming the whole batch. Thus, this model achieves low latency. Apache Flink~\cite{flink} implements the pipelined stream processing model to support a truly native stream processing engine. \projecttitle can adopt this computational model easily by sampling the input data stream in an online manner.

Note that both stream processing models support the sliding window computation~\cite{slider}. The processing window slides over the input stream, whereby the newly incoming data items are added to the window and the old data items are removed from the window. The number of data items within a sliding window may vary in accordance to the arrival rate of data items.

\subsection{Design Assumptions} 
\label{subsec:assumptions}

\projecttitle is based on the following assumptions.  We discuss the possible means to reduce these assumptions in~\S\ref{sec:discussion}. 

\begin{itemize}
\item[1.] We assume there exists a virtual cost function which translates a given query budget (such as the expected latency or throughput guarantees, or the required accuracy level of query results) into the appropriate sample size.

\item[2.] We assume that the input stream is stratified based on the source of data items, i.e., the data items from each sub-stream follow the same distribution and are mutually independent. Here, a \textit{stratum} refers to one sub-stream. If multiple sub-streams have the same distribution, they are combined to form a stratum. 

\item[3.] We assume a time-based window length.  Based on the arrival rate,  the number of data items within a window may vary accordingly. Note that this assumption is consistent with the sliding window APIs in the aforementioned stream processing systems.

\end{itemize}


\subsection{Background: Technical Building Blocks}
\label{sec:preliminaries}

We next describe the two main technical building blocks of \projecttitle: (a) reservoir sampling, and (b) stratified sampling. 

\begin{algorithm}[t]

\myfontsize
\SetLine

\textbf{Input}:  $N$ $\leftarrow$ {\em sample size}\\  

\Begin{
$reservoir$ $\leftarrow$ $\emptyset$; {\commentfontsize // {\em Set of items sampled from the input stream}}\\				
	\ForEach{arriving item $x_{i}$} {
		\If{$|reservoir|$  $<$ $N$}{
		    {\commentfontsize // {\em Fill up the reservoir}}\\
			$reservoir$.{\tt append}($x_{i}$)\;
		} \Else{
		    $p$ $\leftarrow$ $\dfrac{N}{i}$\;
		    {\commentfontsize // {\em Flip a coin comes heads with probability $p$}}\\
		    $head$ $\leftarrow$ {\tt flipCoin}($p$)\;
		    \If {$head$}{
		        {\commentfontsize // {\em Get a random index in the reservoir}}\\
		        $j$ $\leftarrow$ {\tt getRandomIndex}($0$, $|reservoir| - 1$)\;
		        {\commentfontsize // {\em Replace old item in reservoir by $x_{i}$}}\\
		        $reservoir[j]$ $\leftarrow$ $x_{i}$
		    }
		}	    
	}    
}
\caption{\bf  Reservoir algorithm}
\label{alg:reservoirAlgo}
\end{algorithm}

\myparagraph{Reservoir sampling}
\label{subsec:rs}
Suppose we have a stream of data items, and want to randomly select a sample of $N$ items from the stream.  If we know the total number of items in the stream, then the solution is straightforward by applying the simple random sampling~\cite{sampling-3}. 
However, if a stream consists of an unknown number of items or the stream contains a large number of items which could not fit into the storage, then the simple random sampling does not work and a sampling technique called \emph{reservoir sampling} can be used~\cite{reservoir}.

Reservoir sampling receives data items from a stream, and maintains a sample in a buffer called \emph{reservoir}.  Specifically, the technique populates the reservoir with the first $N$ items received from the stream.  After the first $N$ items, every time we receive the $i$-th item ($i > N$), we replace each of the $N$ existing items in the reservoir with the probability of $1/i$, respectively.  In other words, we accept the $i$-th item with the probability of $N/i$, and then randomly replace one existing item in the reservoir.  In doing so, we do not need to know the total number of items in the stream, and reservoir sampling ensures that each item in the stream has equal probability of being selected for the reservoir.  Reservoir sampling is resource-friendly, and its pseudo-code can be found in Algorithm~\ref{alg:reservoirAlgo}.

\begin{figure}[t]
\centering
\includegraphics[scale=0.32]{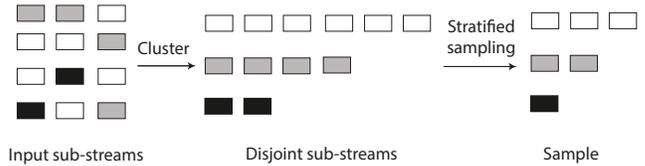}
\caption{Stratified sampling with the sampling fraction of $50\%$.}
\label{fig:stratified-sampling}
\end{figure}
\myparagraph{Stratified sampling}
\label{subsec:sts}
Although reservoir sampling is widely used in stream processing, it could potentially mutilate the statistical quality of the sampled data in the case where the input data stream contains multiple sub-streams with different distributions. This is because reservoir sampling may overlook some sub-streams consisting of only a few data items.   In particular, reservoir sampling does not guarantee that each sub-stream is considered fairly to have its data items selected for the sample.  \emph{Stratified sampling}~\cite{stratified-sampling} 
was proposed to cope with this problem.  Stratified sampling first clusters the input data stream into disjoint sub-streams, and then performs the sampling (e.g., simple random sampling) over each sub-stream independently, as illustrated in Figure~\ref{fig:stratified-sampling}. 
Stratified sampling guarantees that data items from every sub-stream can be fairly selected and no sub-stream will be overlooked.  Stratified sampling, however, works only in the scenario where it knows the statistics of all sub-streams in advance (e.g., the length of each sub-stream).

\section{Design}
\label{sec:design}



In this section, we first present the \projecttitle's workflow (\S\ref{subsec:algo-overview}). Then, we detail its sampling mechanism (\S\ref{subsec:oasrs}), and  its error estimation mechanism (\S\ref{subsec:errEstimation}).

\subsection{System Workflow}
\label{subsec:algo-overview}
Algorithm~\ref{alg:overviewAlgo} presents the workflow of \projecttitle. The algorithm takes the user-specified streaming {\em query} and the query {\em budget} as the input. The algorithm executes the query on the input data stream as a sliding window computation (see \S\ref{subsec:system-model}). 

For each time interval, we first derive the sample size ({\em sampleSize}) using a cost function based on the given query budget (see $\S$\ref{sec:discussion}).  As described in \S\ref{subsec:assumptions}, we currently assume that there exists a cost function which translates a given query budget (such as the expected latency/throughput guarantees, or the required accuracy level of query results) into the appropriate sample size. We discuss the possible means to implement such a cost function in $\S$\ref{sec:discussion}.

We next propose a sampling algorithm (detailed in $\S$\ref{subsec:oasrs}) to select the appropriate $sample$ in an online fashion. Our sampling algorithm further ensures that data items from all sub-streams are fairly selected for the sample, and no single sub-stream is overlooked. 

Thereafter, we execute a data-parallel job to process the user-defined {\em query} on the selected sample. As the last step, we run an error estimation mechanism (as described in $\S$\ref{subsec:errEstimation})  to compute the error bounds for the approximate query result in the form of $output \pm error$ bound.

The whole process repeats for each time interval  as the computation window slides~\cite{tr}. Note that, the query budget can change across time intervals to adapt to user's requirements for the query budget.
\begin{algorithm}[t]

\myfontsize
\SetLine

\textbf{User input}: streaming {\em query} and query {\em budget}\\  

\Begin{
	  {\commentfontsize // {\em Computation in sliding window model ($\S$\ref{subsec:system-model})}}\\
	 \ForEach{time interval} {
		
    {\commentfontsize // {\em Cost function gives the sample size based on the budget  ($\S$\ref{sec:discussion})}}\\
    $sampleSize$ $\leftarrow$ {\tt costFunction}({\em budget})\; 
		
	\ForAll{arriving items in the time interval} {
		 
        {\commentfontsize // {\em Perform OASRS Sampling ($\S$\ref{subsec:oasrs})}}\\
        {\commentfontsize // {\em $W$ denotes the weights of the $sample$}}\\		 	
		$sample$, $W$ $\leftarrow$ {\tt OASRS}($items$, $sampleSize$)\;

	}    
	
	{\commentfontsize // {\em Run query as a data-parallel job to process the sample }}\\
    $output$ $\leftarrow$ {\tt  runJob}($query$, $sample$, $W$)\; 

    {\commentfontsize // {\em Estimate the error bounds of query result/output ($\S$\ref{subsec:errEstimation})}}\\
     $output \pm error$ $\leftarrow$ {\tt estimateError}($output$)\;
    }
}

\caption{\bf: \projecttitle's algorithm overview}
\label{alg:overviewAlgo}
\end{algorithm}



%
%
%

\subsection{Online Adaptive Stratified Reservoir Sampling}
\label{subsec:oasrs}

To realize the real-time stream analytics, we propose a novel sampling technique called Online Adaptive Stratified Reservoir Sampling (OASRS). It achieves both stratified and reservoir samplings without their drawbacks.  Specifically, OASRS does not overlook any sub-streams regardless of their popularity, does not need to know the statistics of sub-streams before the sampling process, and runs efficiently in real time in a distributed manner.



\begin{figure}[t]
\centering
\includegraphics[scale=0.32]{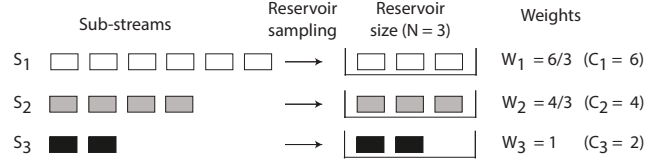}
\caption{OASRS with the reservoirs of size three.}
\label{fig:oasrs-sampling}
\end{figure}

The high-level idea of OASRS is simple, as described in Algorithm~\ref{alg:oasrsAlgo}.  We stratify the input stream into sub-streams according to their sources.  We assume data items from each sub-stream follow the same distribution and are mutually independent. (Here a \textit{stratum} refers to one sub-stream. If multiple sub-streams have the same distribution, they are combined to form a stratum.) We then sample each sub-stream independently, and perform the reservoir sampling for each sub-stream individually.  To do so, every time we encounter a new sub-stream $S_i$, we determine its sample size $N_i$ according to an adaptive cost function considering the specified query budget (see \S\ref{sec:discussion}).  For each sub-stream $S_i$, we perform the traditional reservoir sampling to select items at random from this sub-stream, and ensure that the total number of selected items from $S_i$ does not exceed its sample size $N_i$.  In addition, we maintain a counter $C_i$ to measure the number of items received from $S_i$ within the concerned time interval (see Figure~\ref{fig:oasrs-sampling}).

Applying reservoir sampling to each sub-stream $S_i$ ensures that we can randomly select at most $N_i$ items from each sub-stream.  The selected items from different sub-streams, however, should \emph{not} be treated equally.  In particular, for a sub-stream $S_i$, if $C_i > N_i$ (i.e., the sub-stream $S_i$ has more than $N_i$ items in total during the concerned time interval), then we randomly select $N_i$ items from this sub-stream and each selected item represents $C_i / N_i$ original items on average; otherwise, if $C_i \leq N_i$, we select all the received $C_i$ items so that each selected item only represents itself.  As a result, in order to statistically recreate the original items from the selected items, we assign a specific weight $W_i$ to the items selected from each sub-stream $S_i$:
\begin{equation}
W_i = \Bigg\{
\begin{aligned}
C_i / N_i & \textrm{\qquad if $C_i > N_i$} \\
1\textrm{\qquad} & \textrm{\qquad if $C_i \leq N_i$}
\end{aligned}
\label{eqn:weight}
\end{equation}

We support \emph{approximate linear queries} which return an approximate weighted sum of all items received from all sub-streams.  One example of linear queries is to compute the sum of all received items.  Suppose there are in total $X$ sub-streams $\{S_i\}_{i=1}^{X}$, and from each sub-stream $S_i$ we randomly select at most $N_i$ items.  Specifically, we select $Y_i$ items $\{I_{i,j}\}_{j=1}^{Y_i}$ from each sub-stream $S_i$, where $Y_i \leq N_i$.  In addition, each sub-stream associates with a weight $W_i$ generated according to expression~\ref{eqn:weight}. Then, the approximate sum $SUM_i$ of all items received from each sub-stream $S_i$ can be estimated as:
\begin{equation}
SUM_i = (\sum_{j=1}^{Y_i} I_{i,j}) \times W_i
\end{equation}

As a result, the approximate total sum of all items received from all sub-streams is:
\begin{equation}
SUM = \sum_{i=1}^{X} SUM_i
\end{equation}

A simple extension also enables us to compute the approximate mean value of all received items:
\begin{equation}
MEAN = \frac{SUM}{\sum_{i=1}^{X} C_i}
\label{eqn:mean}
\end{equation}

Here, $C_i$ denotes a counter measuring the number of items received from each sub-stream $S_i$.  Using a similar technique, our OASRS sampling algorithm supports any types of approximate linear queries. This type of queries covers a range of common aggregation queries including, for instance, sum, average, count, histogram, etc. Though linear queries are simple, they can be extended to support a large range of statistical learning algorithms~\cite{BlumDMN05,BlumLR08}.  It is also worth mentioning that, OASRS not only works for a concerned time interval (e.g., a sliding time window), but also works across the entire life cycle of the input data stream.

To summarize, our proposed sampling algorithm combines the benefits of stratified and reservoir samplings via performing the reservoir sampling for each sub-stream  (i.e., stratum) individually.  In addition, our algorithm is an online algorithm since it can perform the ``on-the-fly'' sampling on the input stream without knowing all data items in a window from the beginning~\cite{online-algorithm}.


\myparagraph{Distributed execution}  OASRS can run in a distributed fashion naturally as it does not require synchronization.  One straightforward approach is to make each sub-stream $S_i$ be handled by a set of $w$ worker nodes.  Each worker node samples an equal portion of items from this sub-stream and generates a local reservoir of size no larger than $N_i / w$.  In addition, each worker node maintains a local counter to measure the number of its received items within a concerned time interval for weight calculation.  The rest of the design remains the same.

\begin{algorithm}[t]

\myfontsize
\SetLine

\underline{{\bf {\tt  OASRS($items$, $sampleSize$)}}}\\
\Begin{
    $sample$ $\leftarrow$ $\emptyset$; {\commentfontsize // {\em Set of items sampled within the time interval}}\\
    $S$ $\leftarrow$ $\emptyset$; {\commentfontsize // {\em Set of sub-streams seen so far within the time interval}}\\
    $W$ $\leftarrow$ $\emptyset$; {\commentfontsize // {\em Set of weights of sub-streams within the time interval}}\\
	{\tt Update}($S$);  {\commentfontsize // {\em Update the set of sub-streams}}\\
	{\commentfontsize // {\em Determine the sample size for each sub-stream}}\\
    $N$ $\leftarrow$ {\tt getSampleSize}({\em sampleSize}, S)\; 
	\ForAll{$S_{i}$ in $S$}{
	   	$C_{i}$ $\leftarrow$ $0$; {\commentfontsize // {\em Initial counter to measure \#items in each sub-stream}}\\
		\ForAll{arriving  items in each time interval} {
			{\tt Update}($C_{i}$);  {\commentfontsize // {\em Update the counter}}\\				 	
			$sample_i$ $\leftarrow$ {\tt RS}($items$, $N_i$);  {\commentfontsize // {\em Reservoir sampling }}\\			
			$sample$.{\tt add}($sample_i$); {\commentfontsize // {\em Update the global sample }}\\
			{\commentfontsize // {\em Compute the weight of $sample_i$ according to Equation \ref{eqn:weight}}}\\	
			\If{$C_{i}$  $>$ $N_i$}{
				$W_i$ $\leftarrow$ $\frac{C_{i}}{N_i}$\;   
			}
			\Else{ $W_i$ $\leftarrow$ $1$\; }
			$W$.{\tt add}($W_i$); {\commentfontsize // {\em Update the set of weights}}\\	 
		}
	}    
	
	\Return $sample$, $W$
	
}

\caption{\bf:  Online adaptive stratified reservoir sampling}
\label{alg:oasrsAlgo}
\end{algorithm}

\subsection{Error Estimation}
\label{subsec:errEstimation}

We described how we apply OASRS to randomly sample the input data stream to generate the approximate results for linear queries.  We now describe a method to estimate the accuracy of our approximate results via rigorous error bounds.

Similar to \S\ref{subsec:oasrs}, suppose the input data stream contains $X$ sub-streams $\{S_i\}_{i=1}^{X}$. We compute the approximate sum of all items received from all sub-streams by randomly sampling only $Y_i$ items from each sub-stream $S_i$.  As each sub-stream is sampled independently, the variance of the approximate sum is:
\begin{equation}
Var(SUM) = \sum_{i=1}^{X} Var(SUM_i)
\end{equation}

Further, as items are randomly selected for a sample within each sub-stream, according to the random sampling theory~\cite{samplingBySteve}, the variance of the approximate sum can be estimated as:
\begin{equation}\label{eq:variance_sum}
\widehat{Var}(SUM) = \sum_{i = 1}^{X} \Big(C_i\times(C_i - Y_i)\times \frac{s^2_{i}}{Y_{i}} \Big)
\end{equation}

Here, $C_i$ denotes the total number of items from the sub-stream $S_i$, and $s_i$ denotes the standard deviation of the sub-stream $S_i$'s sampled items:
\begin{equation}
s^2_{i} = \frac{1}{Y_i - 1} \times \sum_{j = 1}^{Y_i} (I_{i,j} - \bar{I_i})^2 \textrm{, where } \bar{I_i} = \frac{1}{Y_i}\times \sum_{j = 1}^{Y_i} I_{i,j}
\end{equation}

Next, we show how we can also estimate the variance of the approximate mean value of all items received from all the $X$ sub-streams.
According to equation~\ref{eqn:mean}, this approximate mean value can be computed as:
\begin{equation}
\begin{split}
MEAN &= \frac{SUM}{\sum_{i=1}^{X} C_i} = \frac{\sum_{i=1}^{X} (C_i \times MEAN_i)}{\sum_{i=1}^{X} C_i} \\
&= \sum_{i=1}^{X} (\omega_i \times MEAN_i)
\end{split}
\end{equation}

Here, $\omega_i = \frac{C_i}{\sum_{i=1}^{X} C_i}$.  Then, as each sub-stream is sampled independently, according to the random sampling theory~\cite{samplingBySteve}, the variance of the approximate mean value can be estimated as:
\begin{equation}\label{eq:variance_mean}
\begin{split}
\widehat{Var}(MEAN)  &=  \sum_{i=1}^{X} Var(\omega_{i} \times MEAN_i) \\
&=  \sum_{i=1}^{X} \Big( \omega_{i}^2 \times Var(MEAN_i) \Big)\\
&=  \sum_{i=1}^{X} \Big( \omega_{i}^2 \times \frac{s^2_{i}}{Y_i}\times \frac{C_i - Y_i}{C_i} \Big)
\end{split}
\end{equation}

Above, we have shown how to estimate the variances of the approximate sum and the approximate mean of the input data stream.  Similarly, by applying the random sampling theory, we can easily estimate the variance of the approximate results of any linear queries.

\myparagraph{Error bound} According to the ``68-95-99.7'' rule~\cite{68-95-99-rule}, our approximate result falls within one, two, and three standard deviations away from the true result with probabilities of 68\%, 95\%, and 99.7\%, respectively, where the standard deviation is the square root of the variance as computed above.
This error estimation is critical because it gives us a quantitative understanding
of the accuracy of our sampling technique.

\section{Implementation}
\label{sec:implementation}
To showcase the effectiveness of our algorithm, we provide two implementations of \projecttitle based on two types of stream processing systems ($\S$\ref{subsec:system-model}): {\em (i)} Apache Spark Streaming~\cite{spark-streaming} --- a batched stream processing system, and {\em (ii)} Apache Flink~\cite{flink} --- a pipelined stream processing system. 

Furthermore, we also built an improved baseline (in addition to the native execution) for Apache Spark, which provides  sampling mechanisms for its machine learning library MLib~\cite{mlib}. In particular, we {\em repurposed} the existing sampling modules available in Apache Spark (primarily used for machine learning) to build an approximate computing system for stream analytics. To have a fair comparison, we evaluate our Spark-based \projecttitle with two baselines: the Spark native execution and the improved Spark sampling based approximate computing system. Meanwhile, Apache Flink does not support sampling operations for stream analytics, therefore we compare our Flink-based \projecttitle with the Flink native execution.

In this section, we first present the necessary background on Apache Spark Streaming (and its existing sampling mechanisms) and Apache Flink ($\S$\ref{subsec:spark-background}). Thereafter, we provide the implementation details of our prototypes ($\S$\ref{subsec:streamapprox-implementation-details}).



\subsection{Background} 
\label{subsec:spark-background}

Apache Spark Streaming and Apache Flink both are DAG-based distributed data processing engines. At a high level, both frameworks provide similar dataflow operators (e.g., map, flatmap, reduce, and filter). However, as described in $\S$\ref{subsec:system-model}, at the core, Spark Streaming is a batched stream processing engine, whereas Flink is a pipelined stream processing engine.

\subsubsection{Apache Spark Streaming}
Apache Spark Streaming splits the input data stream into micro-batches, and for each micro-batch a distributed data-parallel job (Spark job) is launched to process the micro-batch. To sample the input data stream, Spark Streaming makes use of RDD-based sampling functions supported by Apache Spark~\cite{spark-nsdi-2012} to take a sample from each micro-batch. These functions can be classified into the following two categories: 1) Simple Random Sampling (SRS) using {\tt sample}, and 2) Stratified Sampling (STS) using {\tt sampleByKey} and {\tt sampleByKeyExact}. 
 

Simple random sampling (SRS) is implemented using a random sort mechanism~\cite{spark-sampling2} which selects a sample of size $k$ from the input data items in two steps. In the first step, Spark assigns a random number in the range of $[0, 1]$ to each input data item to produce a key-value pair. Thereafter, in the next step, Spark sorts all key-value pairs based on their assigned random numbers, and selects $k$ data items with the smallest assigned random numbers. Since sorting ``Big Data'' is expensive, the second step quickly becomes a bottleneck in this sampling algorithm. To mitigate this bottleneck, Spark reduces the number of items before sorting by setting two thresholds, $p$ and $q$, for the assigned random numbers. In particular, Spark discards the data items with the assigned random numbers larger than $q$, and directly selects data items with the assigned numbers smaller than $p$.  For stratified sampling (STS), Spark first clusters the input data items based on a given criterion (e.g., data sources) to create strata using {\tt groupBy(strata)}.  Thereafter, it applies the aforementioned SRS to data items in each stratum.


\begin{figure}[t]
\centering
\includegraphics[width=0.47\textwidth]{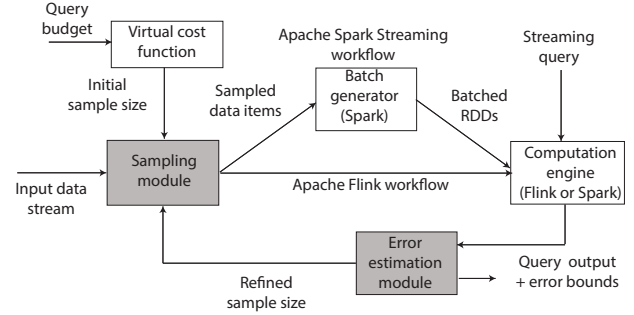}
\caption{Architecture of \projecttitle prototypes (shaded boxes depict the implemented modules). We have implemented our system based on Apache Spark Streaming and Apache Flink.}
\label{fig:streamapprox-implementation}
\end{figure}
\begin{figure*}[t]

\centering

\includegraphics [width=1\textwidth]{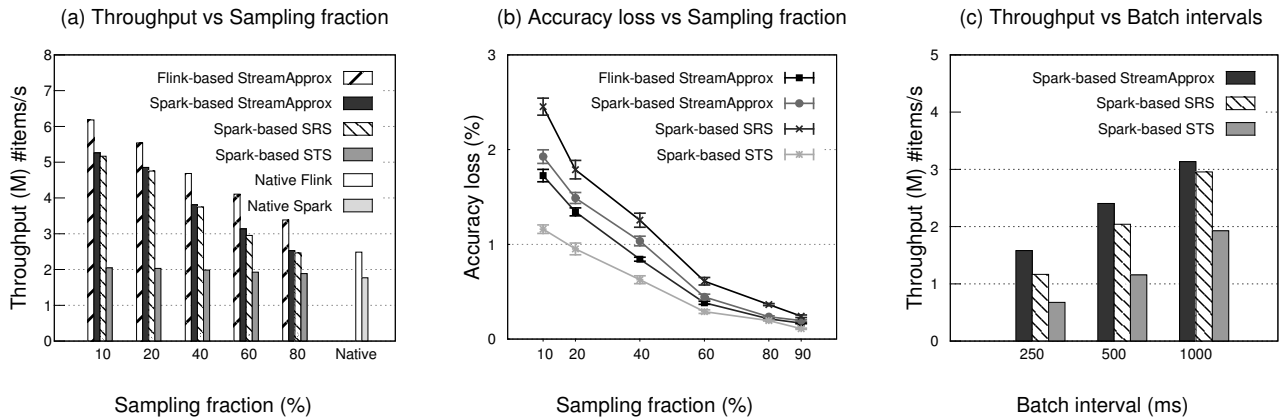}

\caption{Comparison between \projecttitle, Spark-based SRS, Spark-based STS, as well as native Spark and Flink systems.  (a) Peak throughput with varying sampling fractions.  (b) Accuracy loss with varying sampling fractions. (c) Peak throughput with different batch intervals.}


\label{fig:micro-benchmarks-1}

\end{figure*}
\subsubsection{Apache Flink}
In contrast to batched stream processing, Apache Flink adopts a pipelined architecture: whenever an operator in the DAG dataflow emits an item, this item is \emph{immediately} forwarded to the next operator without waiting for a whole data batch. This mechanism makes  Apache Flink a true stream processing engine. In addition, Flink considers batches  as a special case of streaming. Unfortunately, the vanilla Flink does not provide any operations to take a sample of the input data stream. In this work, we provide Flink with an operator to sample input data streams by implementing our proposed sampling algorithm (see \S\ref{subsec:oasrs}).

\subsection{\projecttitle Implementation Details}
\label{subsec:streamapprox-implementation-details}

We next describe the implementation of \projecttitle. Figure~\ref{fig:streamapprox-implementation} illustrates the architecture of our prototypes, where the shaded boxes depict the implemented modules. We showcase workflows for Apache Spark Streaming and Apache Flink in the same figure.

\subsubsection{Spark-based \projecttitle}
In the Spark-based implementation, the input data items are sampled ``on-the-fly'' using our sampling module \emph{before} items are transformed into RDDs. The sampling parameters are determined based on the query budget using a virtual cost function.  In particular, a user can specify the query budget in the form of desired latency or throughput, available computational resources, or acceptable accuracy loss.  As noted in the design assumptions ($\S$\ref{subsec:assumptions}), we have not implemented the virtual cost function since it is beyond the scope of this paper (see \S\ref{sec:discussion} for possible ways to implement such a cost function). Based on the query budget, the virtual cost function determines a sample size, which is then fed to the sampling module.

Thereafter, the sampled input stream is transformed into RDDs, where the data items are split into batches at a pre-defined regular batch interval. Next, the batches are processed as usual using the Spark engine to produce the query output. Since the computed output is an approximate query result, we make use of our error estimation module to give rigorous error bounds. For cases where the error bound is larger than the specified target, an adaptive feedback mechanism is activated to increase the sample size in the sampling module. This way, we achieve higher accuracy in the subsequent epochs.

\myparagraph{I: Sampling module} The sampling module implements the algorithm described in \S\ref{subsec:oasrs} to select samples from the input data stream in an online adaptive fashion.  We modified the Apache Kafka connector of Spark to support our sampling algorithm. In particular, we created a new class {\tt ApproxKafkaRDD} to handle the  input data items from Kafka, which takes required samples to define an RDD for the data items before calling the {\tt compute} function.

\myparagraph{II: Error estimation module} The error estimation module computes the error bounds of the approximate query result. The module also activates a feedback mechanism to re-tune the sample size in the sampling module to achieve the specified accuracy target. We made use of the Apache Common Math library~\cite{math-apache} to implement the error estimation mechanism as described in \S\ref{subsec:errEstimation}.

\subsubsection{Flink-based \projecttitle}


Compared to the Spark-based implementation, implementing a Flink-based \projecttitle is straightforward since Flink supports online stream processing natively. 

\myparagraph{I: Sampling module}  We created a sampling operator by implementing the algorithm described in \S\ref{subsec:oasrs}. This operator samples input data items on-the-fly and in an adaptive manner. The sampling parameters are identified based on the query budget as in Spark-based \projecttitle.

\myparagraph{II: Error estimation module} We reused the  error estimation module implemented in the Spark-based \projecttitle.


\section{Evaluation}
\label{sec:evaluation}

In this section, we present the evaluation results of our implementation. In the next section, we report our experiences on deploying \projecttitle for real-world case studies ($\S$\ref{subsec:case-studies}).


\subsection{Experimental Setup}
\label{subsec:evaluation-setup}

\myparagraph{Synthetic data stream} To understand the effectiveness of our proposed OASRS sampling algorithm, we evaluated \projecttitle using a synthetic input data stream with Gaussian distribution and Poisson distribution. For the Gaussian distribution,  unless specified otherwise, we used  three input sub-streams $A$, $B$, and $C$ with their data items following Gaussian distributions of parameters ($\mu = 10,\  \sigma = 5$),  ($\mu = 1000,\  \sigma = 50$), and  ($\mu = 10000,\  \sigma = 500$), respectively. For the Poisson distribution, unless specified otherwise, we used  three input sub-streams $A$, $B$, and $C$ with their data items following Poisson distributions of parameters ($\lambda = 10$),  ($\lambda = 1000$), and  ($\lambda = 100000000$), respectively.

\myparagraph{Methodology for comparison with Apache Spark} For a fair comparison with the sampling algorithms available in Apache Spark, we also built an Apache Spark-based approximate computing system for stream analytics (as described in $\S$\ref{sec:implementation}). In particular, we used two sampling algorithms available in Spark, namely, Simple Random Sampling (SRS) via {\tt sample}, and Stratified Sampling (STS) via {\tt sampleByKey} and {\tt sampleByKeyExact}.  We applied these sampling operators to each small batch (i.e., RDD) in the input data stream to generate samples.  Note that, the Apache Flink does not support sampling natively.

\myparagraph{Evaluation questions} Our evaluation analyzes the performance of \projecttitle, and compares it with the Spark-based approximate computing system across the following dimensions: (a) varying sample sizes in $\S$\ref{subsec:eval-sample-size}, (b) varying batch intervals in $\S$\ref{subsec:eval-batch-interval}, (c) varying arrival rates for sub-streams in $\S$\ref{subsec:eval-arrival-rate}, (d) varying window sizes in $\S$\ref{subsec:eval-windowsizes}, (e) scalability in $\S$\ref{subsec:scalability}, and (f) skew in the input data stream in $\S$\ref{subsec:eval-skew}. 

%

\subsection{Varying Sample Sizes}
\label{subsec:eval-sample-size}
\begin{figure*}[t]

\centering

\includegraphics [width=1\textwidth]{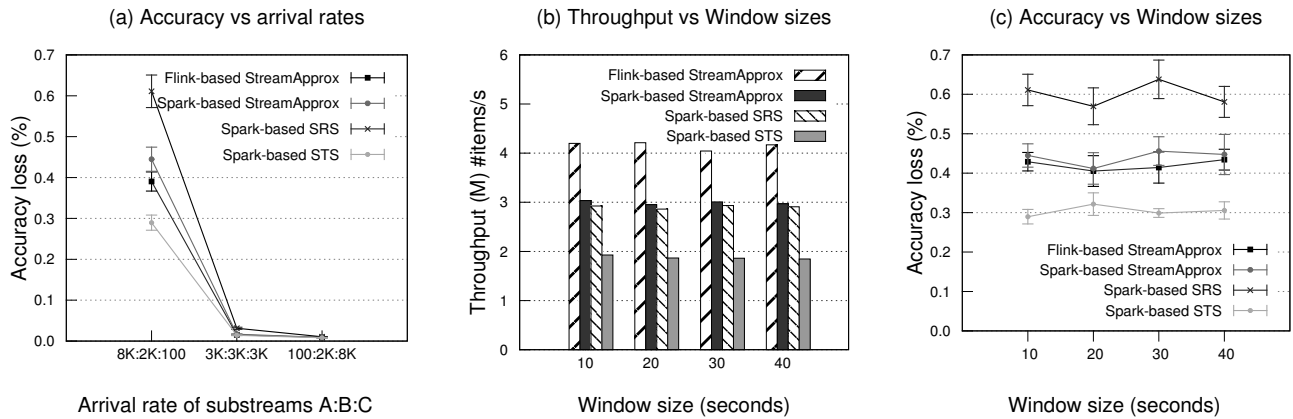}

\caption{Comparison between \projecttitle, Spark-based SRS, and Spark-based STS. (a) Accuracy loss with varying arrival rates.
(b) Peak throughput with varying window sizes. (c) Accuracy loss with varying window sizes. }


\label{fig:micro-benchmarks-2}

\end{figure*}

\myparagraph{Throughput} We first measure the throughput of \projecttitle w.r.t. the Spark- and Flink-based systems with varying sample sizes (sampling fractions). 
To measure the throughput of the evaluated systems, we increase the arrival rate of the input stream until these systems are saturated.

Figure~\ref{fig:micro-benchmarks-1} (a) first shows the throughput comparison of \projecttitle  and the two sampling algorithms in Spark. Spark-based stratified sampling (STS) scales poorly because of its synchronization among Spark workers and the expensive sorting during its sampling process (as detailed in $\S$\ref{subsec:spark-background}). Spark-based \projecttitle achieves a throughput of  $1.68\times$ and $2.60\times$ higher than Spark-based STS with sampling fractions of $60\%$ and $10\%$, respectively. In addition, the Spark-based simple random sampling (SRS) scales better than STS and has a similar throughput as in \projecttitle, but SRS looses the capability of considering each sub-stream fairly.

Meanwhile, Flink-based \projecttitle achieves a throughput of $2.13\times$ and $3\times$ higher than Spark-based STS with sampling fractions of $60\%$ and $10\%$, respectively.  This is mainly due to the fact that Flink is a truly pipelined stream processing engine.  Moreover,  Flink-based \projecttitle achieves a throughput of $1.3\times$ compared to Spark-based \projecttitle and Spark-based SRS with the sampling fraction of $60\%$.

We also compare \projecttitle with native Spark and Flink systems, i.e., without any sampling.  With the sampling fraction of  $60\%$, the throughput of Spark-based \projecttitle is $1.8\times$ higher than the native Spark execution, whereas the throughput of Flink-based \projecttitle is $1.65\times$ higher than the native Flink execution.

\myparagraph{Accuracy}  Next, we compare the accuracy of our proposed OASRS sampling with that of the two sampling mechanisms with the varying sampling fractions.
Figure~\ref{fig:micro-benchmarks-1} (b) first shows that \projecttitle systems and Spark-based STS achieve a higher accuracy than Spark-based SRS. For instance, with the sampling fraction of $60\%$, Flink-based \projecttitle, Spark-based \projecttitle, and Spark-based STS achieve the accuracy loss of $0.38\%$, $0.44\%$, and $0.29\%$, respectively, which are higher than Spark-based SRS which only achieves the accuracy loss of $0.61\%$.  This higher accuracy is due to the fact that both \projecttitle and Spark-based STS integrate stratified sampling which ensures that data items from each sub-stream are selected fairly.
In addition, Spark-based STS achieves even higher accuracy than \projecttitle, but recall that Spark-based STS needs to maintain a sample size of each sub-stream proportional to the size of the sub-stream (see \S\ref{subsec:spark-background}).  This leads to a much lower throughput than \projecttitle which only maintains a sample of a fixed size for each sub-stream.

\subsection{Varying Batch Intervals}
\label{subsec:eval-batch-interval}

Spark-based systems adopt the batched stream processing model.  Next, we evaluate the impact of varying batch intervals on the performance of Spark-based \projecttitle, Spark-based SRS, and Spark-based STS system. We keep the sampling fraction as $60\%$ and measure the throughput of each system with different batch intervals.

Figure~\ref{fig:micro-benchmarks-1} (c) shows that, as the batch interval becomes smaller, the throughput ratio between Spark-based systems gets bigger.  For instance, with the $1000$ms batch interval, the throughput of Spark-based \projecttitle is $1.07\times$ and $1.63\times$ higher than the throughput of Spark-based SRS and STS, respectively; with the $250$ms batch interval, the throughput of \projecttitle is $1.36\times$ and $2.33\times$ higher than the throughput of Spark-based SRS and STS, respectively.  This is because Spark-based \projecttitle samples the data items without synchronization before forming RDDs and significantly reduces costs in scheduling and processing the RDDs, especially when the batch interval is small (i.e., low-latency real-time analytics).

\subsection{Varying Arrival Rates for Sub-Streams}
\label{subsec:eval-arrival-rate}

In the following experiment, we evaluate the impact of varying rates of sub-streams. We used an input data stream with Gaussian distributions as described in  $\S$\ref{subsec:evaluation-setup}.  We maintain the sampling fraction of $60$\% and measure the accuracy loss of the four Spark- and Flink-based systems with different settings of arrival rates.

Figure~\ref{fig:micro-benchmarks-2} (a) shows the accuracy loss of these four systems. The accuracy loss decreases proportionally to the increase of the arrival rate of the sub-stream $C$ which carries the most significant data items compared to other sub-streams. When the arrival rate of the sub-stream $C$ is set to $100$ items/second, Spark-based SRS system achieves the worst accuracy since it may overlook sub-stream $C$ which contributes only a few data items but has significant values. On the other hand, when the arrival rate of sub-stream $C$ is set to $8000$ items/second, the four systems achieve almost the same accuracy. This is mainly because all four systems do not overlook sub-stream $C$ which contains items with the most significant values.

\subsection{Varying Window Sizes}
\label{subsec:eval-windowsizes}

Next, we evaluate the impact of varying window sizes on the throughput and accuracy of the four systems. We used the same input as described in  $\S$\ref{subsec:eval-arrival-rate} with its three sub-streams' arrival rates being 8000, 2000, and 100 items per second. Figure~\ref{fig:micro-benchmarks-2} (b) and Figure~\ref{fig:micro-benchmarks-2} (c) show that the window sizes of the computation do not affect the throughput and accuracy of these systems significantly. This is because the sampling operations are performed at every batch interval in the Spark-based systems and at every slide window interval in the Flink-based \projecttitle.


\subsection{Scalability}
\label{subsec:scalability}
\begin{figure*}[t]

\centering

\includegraphics [width=1\textwidth]{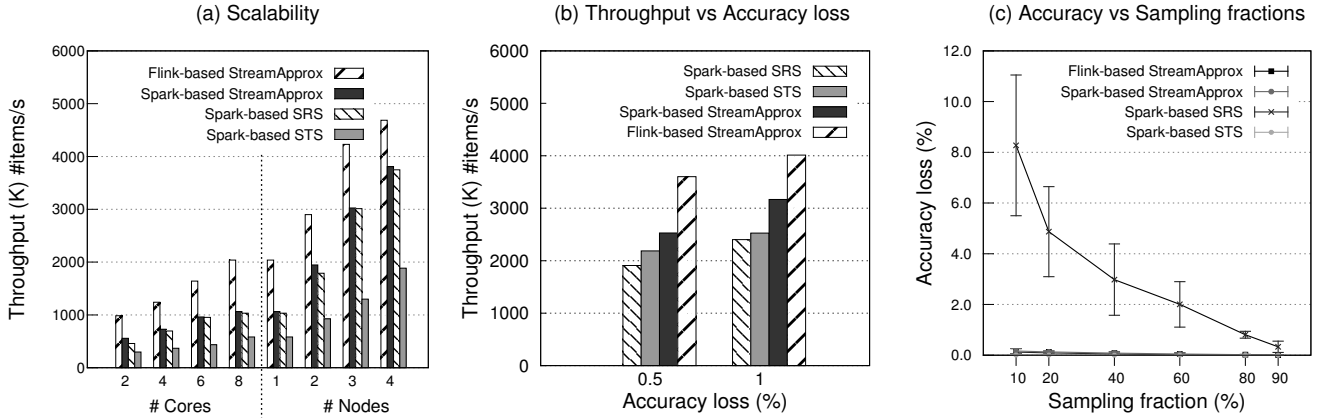}

\caption{Comparison between \projecttitle, Spark-based SRS, and Spark-based STS.  (a) Peak throughput with different numbers of CPU cores and nodes. (b) Peak throughput with accuracy loss. (c) Accuracy loss with varying sampling fractions.}

\vspace{-5mm}

\label{fig:micro-benchmarks-3}

\end{figure*}
\begin{figure*}[t]
\centering
\includegraphics [width=1\textwidth]{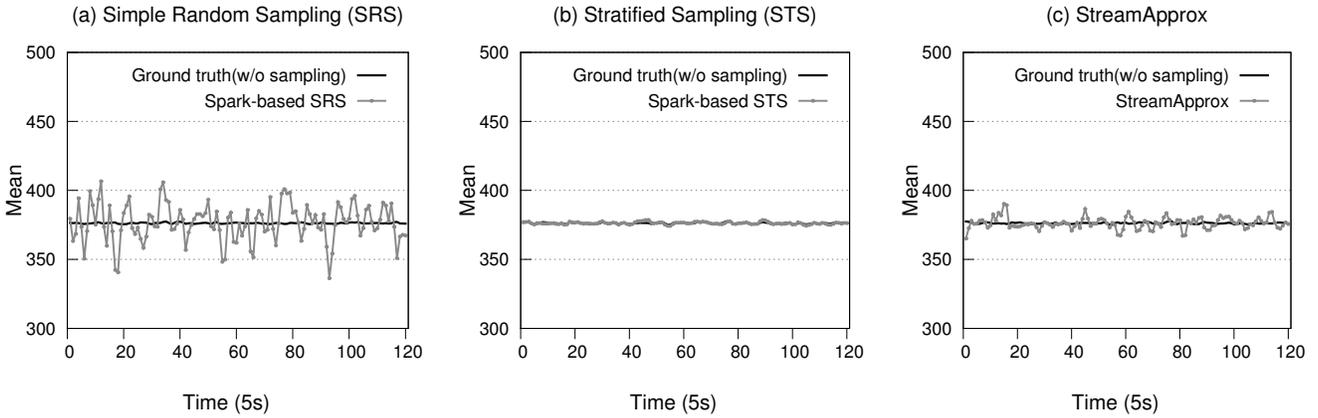}
\caption{The mean values of the received data items produced by different sampling techniques every 5 seconds during a 10-minute observation. The sliding window size is 10 seconds, and each sliding step is 5 seconds.}
\label{fig:sampling-techniques-comparison}
\end{figure*}

To evaluate the scalability of \projecttitle, we keep the sampling fraction as $40\%$ and measure the throughput of \projecttitle and the Spark-based systems with different numbers of CPU cores (scale-up) and different numbers of nodes (scale-out). 

Figure~\ref{fig:micro-benchmarks-3} (a) shows unsurprisingly that \projecttitle and Spark-based SRS scale better than Spark-based STS.  For instance, with one node (8 cores), the throughput of Spark-based \projecttitle and  Spark-based SRS is roughly $1.8\times$ higher than that of Spark-based STS.  With three nodes, Spark-based \projecttitle and  Spark-based SRS achieve a speedup of $2.3\times$ over Spark-based STS. In addition, Flink-based \projecttitle achieves a throughput even $1.9\times$ and $1.4\times$ higher compared to Spark-based \projecttitle with one node and three nodes, respectively.

\subsection{Skew in the Data Stream}
\label{subsec:eval-skew}
\begin{figure*}[ht]
\centering
\includegraphics [width=1\textwidth]{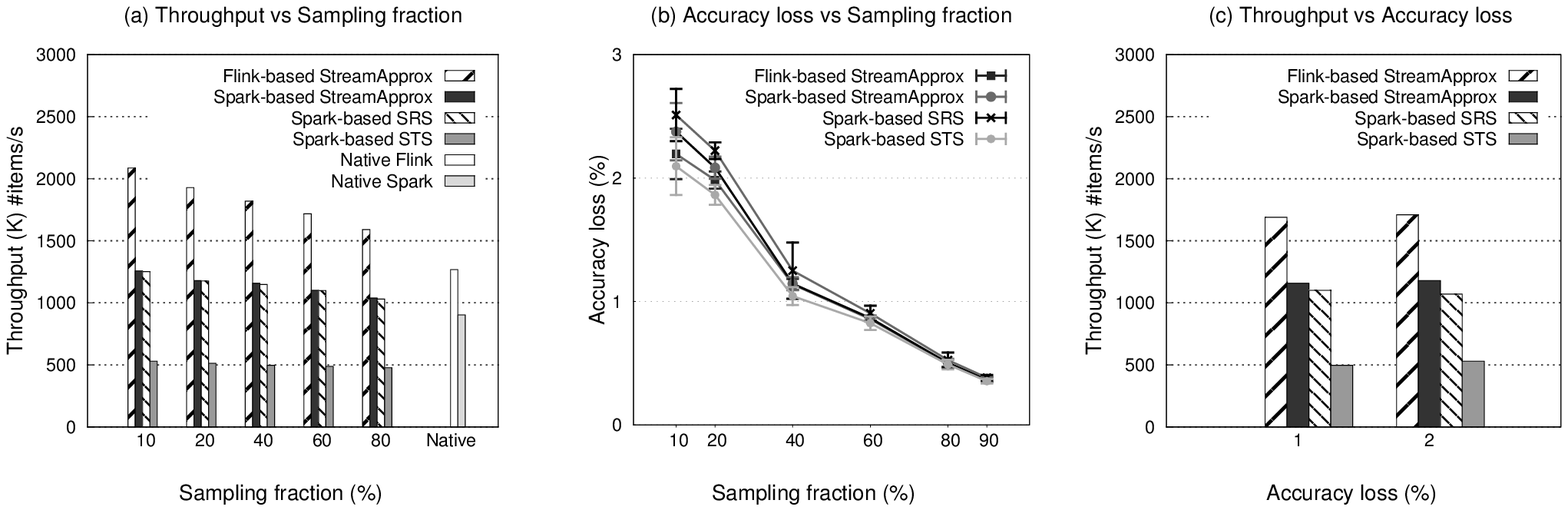}
\caption{The case study of network traffic analytics with a comparison between \projecttitle, Spark-based SRS, Spark-based STS, as well as the native Spark and Flink systems. (a) Peak throughput  with varying sampling fractions. (b) Accuracy loss with varying sampling fractions. (c) Peak throughput with different accuracy losses.}
\label{fig:case-study-network}
\end{figure*}

Lastly, we study the effect of the non-uniformity in sub-stream sizes. In this experiment, we construct an input data stream where one of its sub-streams dominates the other sub-streams. In particular, we evaluated the skew in the input data stream using the following two data distributions:  \emph{(i)} Gaussian distribution and \emph{(ii)} Poisson distribution.

\myparagraph{I: Gaussian distribution} First, we generated an input data stream consisting of three sub-streams $A$, $B$, and $C$ with the Gaussian distribution of parameters ($\mu = 100,\   \sigma = 10$),  ($\mu = 1000,\   \sigma = 100$), and  ($\mu = 10000,\   \sigma = 1000$), respectively. The sub-stream $A$ comprises $80$\% of the data items in the entire data stream, whereas the sub-streams $B$ and $C$ comprise only $19$\% and $1$\% of data items, respectively.  We set the sliding window size to $w = 10$ seconds, and each sliding step to $\delta = 5$ seconds.


Figure~\ref{fig:sampling-techniques-comparison} (a), (b), and (c) present the mean values of the received data items produced by the three Spark-based systems every $5$ seconds during a $10$-minute observation. As expected, Spark-based STS and \projecttitle provide more accurate results than Spark-based SRS because Spark-based STS and \projecttitle ensure that the data items from the minority (i.e., sub-stream $C$) are fairly selected in the samples.

In addition, we keep the accuracy loss across all four systems the same and then measure their respective throughputs. Figure~\ref{fig:micro-benchmarks-3} (b) shows that, with the same accuracy loss of $1\%$, the throughput of Spark-based STS is $1.05\times$ higher than Spark-based SRS, whereas Spark-based \projecttitle achieves a throughput $1.25\times$ higher than Spark-based STS. In addition, Flink-based \projecttitle achieves the highest throughput which is $1.68\times$, $1.6\times$, and $1.26\times$ higher than Spark-based SRS, Spark-based STS, and Spark-based \projecttitle, respectively.

\myparagraph{II: Poisson distribution}  In the next experiment, we generated an input data stream with the Poisson distribution as described in  $\S$\ref{subsec:evaluation-setup}. The sub-stream $A$  accounts for $80$\% of the entire data stream items, while the sub-stream $B$ accounts for $19.99$\% and the sub-stream $C$ comprises only $0.01$\% of the data stream items, respectively. Figure~\ref{fig:micro-benchmarks-3} (c) shows that \projecttitle  systems and Spark-based STS  outperform Spark-based SRS in terms of accuracy. The reason for this is \projecttitle systems and Spark-based STS do not overlook sub-stream $C$ which has items with significant values. Furthermore, this result strongly demonstrates the superiority of the proposed sampling algorithm OASRS over simple random sampling in processing long-tail data which is very common in practice.

\section{Case Studies}
\label{subsec:case-studies}

In this section, we report our experiences and results with the following two real-world case studies: (a) network traffic analytics ($\S$\ref{subsec:eval-case-study-network}) and (b) New York taxi ride analytics ($\S$\ref{subsec:eval-case-study-taxi}).

\subsection{Experimental Setup}

\myparagraph{Cluster setup} We performed experiments using a cluster of $17$ nodes. Each node in the cluster has $2$ Intel Xeon E5405 CPUs (quad core), $8$GB of RAM, and a SATA-2 hard disk, running Ubuntu $14.04.5$ LTS. We deployed our \projecttitle prototype on $5$ nodes ($1$ driver node and $4$ worker nodes), the traffic replay tool on $5$ nodes, the Apache Kafka-based stream aggregator on $4$ nodes, and the Apache Zookeeper~\cite{zookeeper} on the remaining $3$ nodes.

\myparagraph{Measurements} We evaluated \projecttitle using the following key metrics: (a) throughput: measured as the number of data items processed per second; (b) latency: measured as the total time required for processing the respective dataset; and lastly, (c) accuracy loss: measured as $|approx - exact| / exact$ where $approx$ and $exact$ denote the results from \projecttitle and a native system without sampling, respectively. 

\myparagraph{Methodology} We built a tool to efficiently replay the case-study dataset as the input data stream.  
In particular, for the throughput measurement, 
we tuned the replay tool to first feed $2000$ messages/second and continued to increase the throughput until the system was saturated. Here, each message contained $200$ data items.  

For comparison, we report results from \projecttitle, Spark-based SRS, Spark-based STS systems, as well as the native Spark and Flink systems. For all experiments,  we report measurements based on the average over $10$ runs.  Lastly, the sliding window size was set to $10$ seconds, and each sliding step was set to $5$ seconds.
\begin{figure*}[t]
\centering
\includegraphics [width=1\textwidth]{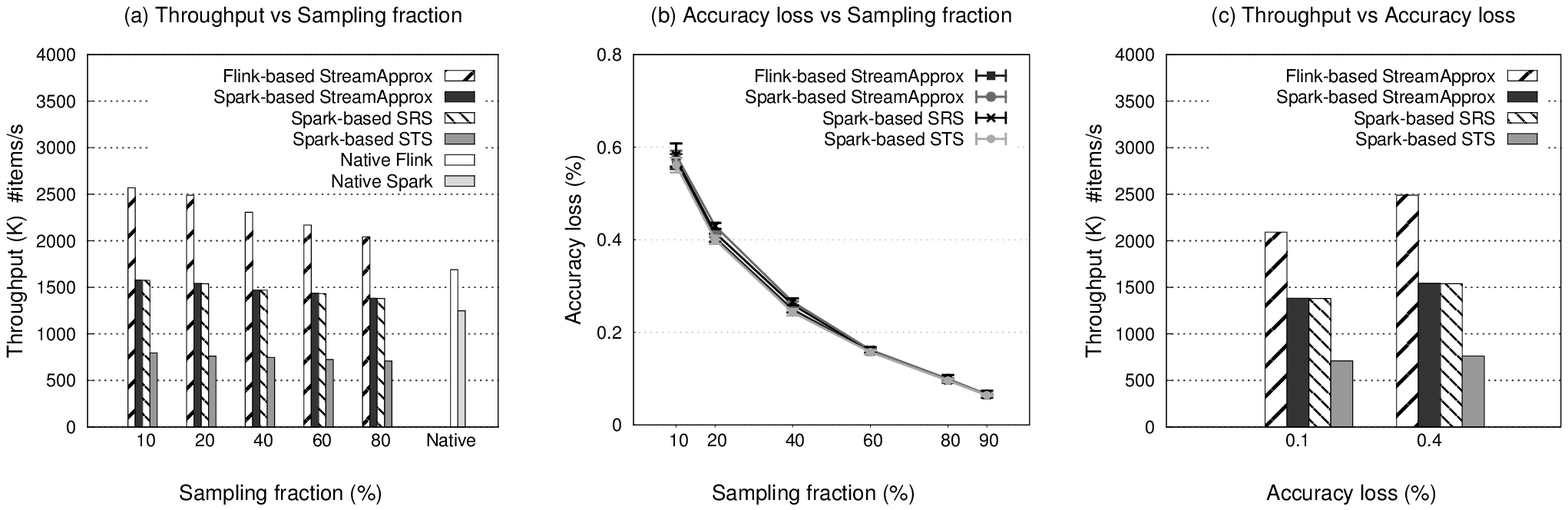}
\caption{The case study of New York taxi ride analytics with a comparison between  \projecttitle, Spark-based SRS, Spark-based STS, as well as the native Spark and Flink systems. (a) Peak throughput with varying sampling fractions. (b) Accuracy loss with varying sampling fractions. (c) Peak throughput with different accuracy losses.}
\label{fig:case-study-taxi}
\end{figure*}

\subsection{Network Traffic Analytics}
\label{subsec:eval-case-study-network}

In the first case study, we deployed \projecttitle for a real-time network traffic monitoring application to measure the TCP, UDP, and ICMP network traffic over time.

\myparagraph{Dataset} We used the publicly-available 670GB network traces from CAIDA~\cite{caida2015}.  These were recorded on the high-speed Internet backbone links in Chicago in 2015.  We converted the raw network traces into the NetFlow format~\cite{netflow}, and then removed unused fields (such as source and destination ports, duration, etc.) in each NetFlow record to build a dataset for our experiments. 

\myparagraph{Query} We deployed the evaluated systems to measure the total sizes of TCP, UDP, and ICMP network traffic in each sliding window.



\myparagraph{Results} Figure~\ref{fig:case-study-network} (a) presents the throughput comparison between \projecttitle, Spark-based SRS, Spark-based STS systems, as well as the native Spark and Flink systems. The result shows that Spark-based \projecttitle achieves more than $2\times$ throughput than Spark-based STS, and achieves a similar throughput compared with Spark-based SRS (which looses the capability of considering each sub-stream fairly). In addition, due to Flink's pipelined stream processing model, Flink-based \projecttitle achieves a throughput even $1.6\times$ higher than Spark-based \projecttitle and Spark-based SRS. We also compare \projecttitle with the native Spark and Flink systems.  With the sampling fraction of  $60\%$, the throughput of Spark-based \projecttitle is $1.3\times$ higher than the native Spark execution, whereas the throughput of Flink-based \projecttitle is $1.35\times$ higher than the native Flink execution. Surprisingly, the throughput of the native Spark execution is even higher than the throughput of Spark-based STS. The reason for this is that  Spark-based STS requires the expensive extra steps (see $\S$\ref{subsec:spark-background}).

Figure~\ref{fig:case-study-network} (b) shows the accuracy loss with different sampling fractions. As the sampling fraction increases, the accuracy loss of \projecttitle, Spark-based SRS, and Spark-based STS decreases (i.e., accuracy improves), but not linearly. \projecttitle systems produce more accurate results than Spark-based SRS but less accurate results than Spark-based STS.  Note however that, although both \projecttitle systems and Spark-based STS integrate stratified sampling to ensure that every sub-stream is considered fairly,  \projecttitle systems are much more resource-friendly than Spark-based STS.  This is because Spark-based STS requires synchronization among workers for the expensive join operation to take samples from the input data stream, whereas \projecttitle performs the sampling operation with a configurable sample size for sub-streams requiring no synchronization between workers.


In addition, to show the benefit of \projecttitle, we fixed the same accuracy loss for all four systems and then compared their respective throughputs. Figure~\ref{fig:case-study-network} (c) shows that, with the accuracy loss of $1\%$, the throughput of Spark-based \projecttitle is $2.36\times$ higher than Spark-based STS, and 1.05$\times$ higher than Spark-based SRS.  Flink-based \projecttitle achieves a throughput even $1.46\times$ higher than Spark-based \projecttitle.

Finally, to make a comparison in terms of latency between these systems, we implemented our proposed sampling algorithm OASRS in Spark-core, and then measured the latency in processing the network traffic dataset.
Figure~\ref{fig:latency-case-studies} indicates that the latency of Spark-based \projecttitle is 1.39$\times$ and 1.69$\times$ lower than Spark-based SRS and Spark-based STS in processing the network traffic dataset.
\subsection{New York Taxi Ride Analytics}
\label{subsec:eval-case-study-taxi}

In the second case study, we evaluated \projecttitle with a taxi ride dataset to measure the average distance of trips starting from different boroughs in New York City.

\myparagraph{Dataset} We used the {\em NYC Taxi Ride} dataset from the DEBS 2015 Grand Challenge~\cite{nyc-taxi-dataset}. The dataset consists of the itinerary information of all rides across $10,000$ taxies in New York City in 2013. In addition, we mapped the start coordinates of each trip in the dataset into one of the six boroughs in New York. 

\myparagraph{Query} We deployed \projecttitle, Spark-based SRS, Spark-based STS systems, as well as the native Spark and Flink systems to measure the average distance of the trips starting from various boroughs in each sliding window.



\myparagraph{Results}
Figure~\ref{fig:case-study-taxi} (a) shows that Spark-based \projecttitle achieves a similar throughput compared with Spark-based SRS (which, however, does not consider each sub-stream fairly), and a roughly $2\times$ higher throughput than Spark-based STS. In addition, due to Flink's pipelined streaming model, Flink-based \projecttitle achieves a $1.5\times$ higher throughput  compared to Spark-based \projecttitle and Spark-based SRS. 
We again compared \projecttitle with the native Spark and Flink systems.
With the sampling fraction of  $60\%$, the throughput of Spark-based \projecttitle is $1.2\times$ higher than the throughput of the native Spark execution, whereas the throughput of Flink-based \projecttitle is $1.28\times$ higher than the throughput of the native Flink execution. Similar to the result in the first case study, the throughput of the native Spark execution is higher than throughput of Spark-based STS.

\begin{figure}[t]
\centering
\includegraphics [width=0.3\textwidth]{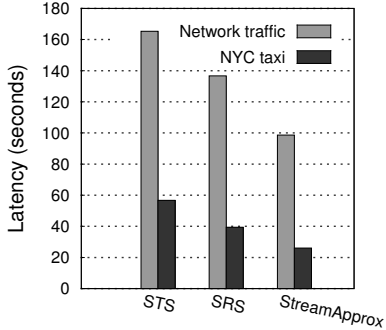}
\caption{The latency comparison between  \projecttitle, Spark-based SRS, and Spark-based STS with the real-world datasets. The sampling fraction is set to 60\%.}
\label{fig:latency-case-studies}
\end{figure}

Figure~\ref{fig:case-study-taxi} (b) depicts the accuracy loss of these systems with different sampling fractions. The results show that they all achieve a very similar accuracy in this case study.
In addition, we also fixed the same accuracy loss of 1\% for all four systems to measure their respective throughputs. Figure~\ref{fig:case-study-taxi} (c) shows that Flink-based \projecttitle achieves the best throughput which is $1.6\times$ higher than Spark-based \projecttitle and Spark-based SRS, and $3\times$ higher than Spark-based STS.
Figure~\ref{fig:latency-case-studies} further indicates that Spark-based \projecttitle provides the 1.52$\times$ and 2.18$\times$ lower latency than Spark-based SRS and Spark-based STS in processing the NYC taxi ride dataset.

\section{Discussion}
\label{sec:discussion}
The design of \projecttitle~is based on the assumptions mentioned in $\S$\ref{subsec:assumptions}. Reducing these assumptions is beyond the scope of this paper.  Nevertheless, in this section, we discuss some approaches that could be used to meet our assumptions.

\myparagraph{I: Virtual cost function} We currently assume that there exists a virtual cost function to translate a user-specified query budget into the sample size. The query budget could be specified, for instance, as either available computing resources, desired accuracy or desired latency requirement. 

For instance, with an accuracy budget, we can define the sample size for each sub-stream based on a desired width of the confidence interval using Equation~\ref{eq:variance_mean} and the ``68-95-99.7'' rule. With a desired latency budget, users can specify it by defining the window time interval or the slide interval for their computations over the input data stream. It becomes a bit more challenging to specify a budget for resource utilization.  Nevertheless, we discuss some existing techniques that could be used to implement such a cost function to achieve the desired resource target. In particular, we refer to the two existing techniques:  (a) Virtual data center~\cite{pulsar}, and (b) resource prediction model~\cite{conductor-nsdi-2012, conductor-podc-2010, conductor-ladis-2010} 
for the latency requirement. 
 
Pulsar~\cite{pulsar} proposes an abstraction of a virtual data center (VDC) to provide performance guarantees to tenants in the cloud. In particular, Pulsar makes use of a virtual cost function to translate the cost of a request processing into the required computational resources using a multi-resource token algorithm.  We could adapt the cost function for our purpose as follows: we consider a data item in the input stream as a request and the ``amount of resources'' required to process it as the cost in tokens. Also, the given resource budget is converted in the form of tokens, using the pre-advertised cost model per resource. This allows us to compute the number of items, i.e., the sample size, that can be processed within the given resource budget. 

For any given latency requirement, we could employ a resource prediction model~\cite{conductor-nsdi-2012}. 
In particular, we could build the prediction model by analyzing the diurnal patterns in resource usage~\cite{googlecluster} to predict the future resource requirement for the given latency budget. This resource requirement can then be mapped to the desired sample size based on the same approach as described above.

\myparagraph{II: Stratified sampling} In our design in $\S$\ref{sec:design}, we currently assume that the input stream is already stratified based on the source of events, i.e., the data items within each stratum follow the same distribution.  This assumption is practical in many cases. For example, consider an IoT use-case which analyzes data streams from sensors to measure the temperature of a city. The data stream from each individual sensor will follow the same distribution since it measures the temperature at the same location in the city. Therefore, a straightforward way to stratify the input data streams is to consider each sensor's data stream as a stratum (sub-stream). In more complex cases when we cannot classify strata based on the sources, we need a preprocessing step to stratify the input data stream. This stratification problem is orthogonal to our work, nevertheless for completeness, we discuss two proposals for the stratification of evolving data streams, namely bootstrap~\cite{bootstrap-Dziuda} and semi-supervised learning~\cite{semi-supervised-algorithm}.

Bootstrap~\cite{bootstrap-Dziuda} 
is a well-studied non-parametric sampling technique in statistics for the estimation of distribution for a given population. In particular, the bootstrap technique randomly selects ``bootstrap samples'' with replacement to estimate the unknown parameters of a population, for instance, by averaging the bootstrap samples. We can employ a bootstrap-based estimator for the stratification of incoming sub-streams. Alternatively, we could also make use of a semi-supervised algorithm~\cite{semi-supervised-algorithm} to stratify a data stream. The advantage of this algorithm is that it can work with both labeled and unlabeled data streams to train a classification model.

\section{Related Work}
\label{sec:related}
Given the advantages of making a trade-off between accuracy and efficiency, 
approximate computing is applied to various domains: graphics, machine learning, scientific simulations, etc. In this context, 
approximation mechanisms have been proposed at various levels of the system stack, from hardware to applications --- including languages, tools, processors, accelerators, memory, and compilers (refer to~\cite{sampson-thesis} for a detailed survey). Our work mainly builds on the advancements in the databases community. In this section, we survey the approximation techniques in this context. 

Over the last two decades, the databases community has proposed various approximation techniques based on sampling~\cite{stratified-sampling, sampling-2},  online aggregation~\cite{online-aggregation}, and sketches~\cite{sketching}.  These techniques make different trade-offs with respect to the output quality, supported query interface, and workload. However, the early work in approximate computing mainly targeted towards the centralized database architecture.

Recently, sampling-based approaches have been successfully adopted for distributed data analytics~\cite{BlinkDB,approxhadoop, quickr-sigmod, incapprox-www-2016}. 
In particular, BlinkDB~\cite{BlinkDB} proposes an approximate distributed query processing engine that uses stratified sampling~\cite{stratified-sampling} to support ad-hoc queries with error and response time constraints.  ApproxHadoop~\cite{approxhadoop} uses multi-stage sampling~\cite{sampling-3} for approximate MapReduce job execution.  Both BlinkDB and ApproxHadoop show that  it is possible to make  a trade-off between the output accuracy and the performance gains (also the efficient resource utilization) by employing sampling-based approaches to compute over a subset of data items.   However, these ``big data'' systems target batch processing and cannot provide required low-latency guarantees for stream analytics.

Like BlinkDB, Quickr~\cite{quickr-sigmod} also supports complex ad-hoc queries in big-data clusters. Quickr deploys distributed sampling operators to reduce execution costs of parallelized queries. In particular, Quickr first injects sampling operators into the query plan; thereafter, it searches for an optimal query plan among sampled query plans to execute input queries. However, Quickr is also designed for static databases, and it does not account for stream analytics. IncApprox~\cite{incapprox-www-2016}  is a data analytics system that combines two computing paradigms together, namely, approximate and incremental computations~\cite{incoop, ithreads, shredder, incoop-hotcloud} for  stream analytics. 
The system is  based on an online ``biased sampling'' algorithm that uses self-adjusting computation~\cite{Bhatotia15} to produce incrementally updated approximate output.
Lastly, PrivApprox~\cite{PrivApprox2017} supports privacy-preserving data analytics using a combination of randomized response and approximate computation. 

By contrast, in \projecttitle, we designed an ``online'' sampling algorithm solely for approximate computing, while avoiding the  limitations of existing sampling algorithms.

\section{Conclusion}
\label{sec:conclusion}

In this paper, we presented \projecttitle, a stream analytics system for approximate computing. \projecttitle allows users to make a systematic trade-off between the output accuracy and the computation efficiency. To achieve this goal, we designed an online stratified reservoir sampling algorithm which ensures the statistical quality of the sample from the input data stream. Our proposed sampling algorithm is generalizable to two prominent types of stream processing models: batched and pipelined stream processing models.  

To showcase the effectiveness of our proposed algorithm,  we built \projecttitle based on Apache Spark Streaming and Apache Flink. We evaluated the effectiveness of our system using a series of micro-benchmarks and real-world case studies. Our evaluation shows that, with varying sampling fractions of $80\%$ to $10\%$, Spark- and Flink-based \projecttitle achieves a significantly higher throughput of $1.15\times$---$3\times$ compared to the native Spark Streaming and Flink executions, respectively. Furthermore, \projecttitle achieves a speedup of $1.1\times$---$2.4\times$ compared to a Spark-based sampling system for approximate computing, while maintaining the same level of accuracy for the query output. 
Finally, the source code of \projecttitle is publicly available: \href{https://streamapprox.github.io/}{https://streamapprox.github.io/}.

\balance
\bibliographystyle{abbrv}
\bibliography{main}  

\end{document}